



%

%
 \font\twelvebf=cmbx12
 \font\twelvett=cmtt12
 \font\twelveit=cmti12
 \font\twelvesl=cmsl12
 \font\twelverm=cmr12		\font\ninerm=cmr9
 \font\twelvei=cmmi12		\font\ninei=cmmi9
 \font\twelvesy=cmsy10 at 12pt	\font\ninesy=cmsy9
 \skewchar\twelvei='177		\skewchar\ninei='177
 \skewchar\seveni='177	 	\skewchar\fivei='177
 \skewchar\twelvesy='60		\skewchar\ninesy='60
 \skewchar\sevensy='60		\skewchar\fivesy='60
%
%

%
 \font\fourteenrm=cmr12 scaled 1200
 \font\seventeenrm=cmr12 scaled 1440
 \font\fourteenbf=cmbx12 scaled 1200
 \font\seventeenbf=cmbx12 scaled 1440
%
%

%
%
%
\font\tenmsb=msbm10
\font\twelvemsb=msbm10 scaled 1200
\newfam\msbfam

%
\font\tensc=cmcsc10
\font\twelvesc=cmcsc10 scaled 1200
\newfam\scfam

%
\def\seventeenpt{\def\rm{\fam0\seventeenrm}%
 \textfont\bffam=\seventeenbf	\def\bf{\fam\bffam\seventeenbf}}
\def\fourteenpt{\def\rm{\fam0\fourteenrm}%
 \textfont\bffam=\fourteenbf	\def\bf{\fam\bffam\fourteenbf}}
\def\twelvept{\def\rm{\fam0\twelverm}%
 \textfont0=\twelverm	\scriptfont0=\ninerm	\scriptscriptfont0=\sevenrm
 \textfont1=\twelvei	\scriptfont1=\ninei	\scriptscriptfont1=\seveni
 \textfont2=\twelvesy	\scriptfont2=\ninesy	\scriptscriptfont2=\sevensy
 \textfont3=\tenex	\scriptfont3=\tenex	\scriptscriptfont3=\tenex
 \textfont\itfam=\twelveit	\def\it{\fam\itfam\twelveit}%
 \textfont\slfam=\twelvesl	\def\sl{\fam\slfam\twelvesl}%
 \textfont\ttfam=\twelvett	\def\tt{\fam\ttfam\twelvett}%
 \scriptfont\bffam=\tenbf 	\scriptscriptfont\bffam=\sevenbf
 \textfont\bffam=\twelvebf	\def\bf{\fam\bffam\twelvebf}%
 \textfont\scfam=\twelvesc	\def\sc{\fam\scfam\twelvesc}%
 \textfont\msbfam=\twelvemsb	
 \baselineskip 14pt%
 \abovedisplayskip 7pt plus 3pt minus 1pt%
 \belowdisplayskip 7pt plus 3pt minus 1pt%
 \abovedisplayshortskip 0pt plus 3pt%
 \belowdisplayshortskip 4pt plus 3pt minus 1pt%
 \parskip 3pt plus 1.5pt
 \setbox\strutbox=\hbox{\vrule height 10pt depth 4pt width 0pt}}
\def\tenpt{\def\rm{\fam0\tenrm}%
 \textfont0=\tenrm	\scriptfont0=\sevenrm	\scriptscriptfont0=\fiverm
 \textfont1=\teni	\scriptfont1=\seveni	\scriptscriptfont1=\fivei
 \textfont2=\tensy	\scriptfont2=\sevensy	\scriptscriptfont2=\fivesy
 \textfont3=\tenex	\scriptfont3=\tenex	\scriptscriptfont3=\tenex
 \textfont\itfam=\tenit		\def\it{\fam\itfam\tenit}%
 \textfont\slfam=\tensl		\def\sl{\fam\slfam\tensl}%
 \textfont\ttfam=\tentt		\def\tt{\fam\ttfam\tentt}%
 \scriptfont\bffam=\sevenbf 	\scriptscriptfont\bffam=\fivebf
 \textfont\bffam=\tenbf		\def\bf{\fam\bffam\tenbf}%
 \textfont\scfam=\tensc		\def\sc{\fam\scfam\tensc}%
 \textfont\msbfam=\tenmsb	
 \baselineskip 12pt%
 \abovedisplayskip 6pt plus 3pt minus 1pt%
 \belowdisplayskip 6pt plus 3pt minus 1pt%
 \abovedisplayshortskip 0pt plus 3pt%
 \belowdisplayshortskip 4pt plus 3pt minus 1pt%
 \parskip 2pt plus 1pt
 \setbox\strutbox=\hbox{\vrule height 8.5pt depth 3.5pt width 0pt}}

%
\def\twelvepoint{%
 \def\small{\tenpt\rm}%
 \def\normal{\twelvept\rm}%
 \def\big{\fourteenpt\rm}%
 \def\huge{\seventeenpt\rm}%
 \footline{\hss\twelverm\folio\hss}
 \normal}
%

%
\def\bigbold{\big\bf}

%
\catcode`\@=11
%
%
\def\footnote#1{\edef\@sf{\spacefactor\the\spacefactor}#1\@sf
 \insert\footins\bgroup\small
 \interlinepenalty100	\let\par=\endgraf
 \leftskip=0pt		\rightskip=0pt
 \splittopskip=10pt plus 1pt minus 1pt	\floatingpenalty=20000
 \smallskip\item{#1}\bgroup\strut\aftergroup\@foot\let\next}
%
%
%
%
\def\hexnumber@#1{\ifcase#1 0\or 1\or 2\or 3\or 4\or 5\or 6\or 7\or 8\or
 9\or A\or B\or C\or D\or E\or F\fi}
\edef\msbfam@{\hexnumber@\msbfam}

%
%
%
\catcode`\@=12

\newcount\EQNO      \EQNO=0
\newcount\FIGNO     \FIGNO=0
\newcount\REFNO     \REFNO=0
\newcount\SECNO     \SECNO=0
\newcount\SUBSECNO  \SUBSECNO=0
\newcount\FOOTNO    \FOOTNO=0
\newbox\FIGBOX      \setbox\FIGBOX=\vbox{}
\newbox\REFBOX      \setbox\REFBOX=\vbox{}
\newbox\RefBoxOne   \setbox\RefBoxOne=\vbox{}

\expandafter\ifx\csname normal\endcsname\relax\def\normal{\null}\fi

\def\Eqno{\global\advance\EQNO by 1 \eqno(\the\EQNO)%
    \gdef\label##1{\xdef##1{\nobreak(\the\EQNO)}}}
\def\Fig#1{\global\advance\FIGNO by 1 Figure~\the\FIGNO%
    \global\setbox\FIGBOX=\vbox{\unvcopy\FIGBOX
      \narrower\smallskip\item{\bf Figure \the\FIGNO~~}#1}}
\def\Ref#1{\global\advance\REFNO by 1 \nobreak[\the\REFNO]%
    \global\setbox\REFBOX=\vbox{\unvcopy\REFBOX\normal
      \smallskip\item{\the\REFNO .~}#1}%
    \gdef\label##1{\xdef##1{\nobreak[\the\REFNO]}}}
\def\Section#1{\SUBSECNO=0\advance\SECNO by 1
    \bigskip\leftline{\bf \the\SECNO .\ #1}\nobreak}
\def\Subsection#1{\advance\SUBSECNO by 1
    \medskip\leftline{\bf \ifcase\SUBSECNO\or
    a\or b\or c\or d\or e\or f\or g\or h\or i\or j\or k\or l\or m\or n\fi
    )\ #1}\nobreak}
\def\Footnote#1{\global\advance\FOOTNO by 1
    \footnote{\nobreak$\>\!{}^{\the\FOOTNO}\>\!$}{#1}}
\def\SameFootnote{$\>\!{}^{\the\FOOTNO}\>\!$}

\def\References{\bigskip\centerline{\bf REFERENCES}
                \smallskip\copy\REFBOX}
\def\NewRefPage{\setbox\RefBoxOne=\vbox{\unvcopy\REFBOX}
		\setbox\REFBOX=\vbox{}
		\def\References{\bigskip\centerline{\bf REFERENCES}
                		\nobreak\smallskip\nobreak\copy\RefBoxOne
				\vfill\eject
				\smallskip\copy\REFBOX}
		\def\NewRefPage{}}




\font\twelvebm=cmmib10 at 12pt
\font\tenbm=cmmib10
\font\ninei=cmmi9
\newfam\bmfam

\def\twelvepointbmit{
\textfont\bmfam=\twelvebm
\scriptfont\bmfam=\ninei
\scriptscriptfont\bmfam=\seveni
\def\bmit{\fam\bmfam\twelvebm}
}

\def\tenpointbmit{
\textfont\bmfam=\tenbm
\scriptfont\bmfam=\seveni
\scriptscriptfont\bmfam=\fivei
\def\bmit{\fam\bmfam\tenbm}
}

\tenpointbmit

\mathchardef\Gamma="7100
\mathchardef\Delta="7101
\mathchardef\Theta="7102
\mathchardef\Lambda="7103
\mathchardef\Xi="7104
\mathchardef\Pi="7105
\mathchardef\Sigma="7106
\mathchardef\Upsilon="7107
\mathchardef\Phi="7108
\mathchardef\Psi="7109
\mathchardef\Omega="710A
\mathchardef\alpha="710B
\mathchardef\beta="710C
\mathchardef\gamma="710D
\mathchardef\delta="710E
\mathchardef\epsilon="710F
\mathchardef\zeta="7110
\mathchardef\eta="7111
\mathchardef\theta="7112
\mathchardef\iota="7113
\mathchardef\kappa="7114
\mathchardef\lambda="7115
\mathchardef\mu="7116
\mathchardef\nu="7117
\mathchardef\xi="7118
\mathchardef\pi="7119
\mathchardef\rho="711A
\mathchardef\sigma="711B
\mathchardef\tau="711C
\mathchardef\upsilon="711D
\mathchardef\phi="711E
\mathchardef\cho="711F
\mathchardef\psi="7120
\mathchardef\omega="7121
\mathchardef\varepsilon="7122
\mathchardef\vartheta="7123
\mathchardef\varpi="7124
\mathchardef\varrho="7125
\mathchardef\varsigma="7126
\mathchardef\varphi="7127



%
%
\twelvepoint			
%
%



\font\twelvebm=cmmib10 at 12pt
\font\tenbm=cmmib10
\font\ninei=cmmi9
\newfam\bmfam

\def\twelvepointbmit{
\textfont\bmfam=\twelvebm
\scriptfont\bmfam=\ninei
\scriptscriptfont\bmfam=\seveni
\def\bmit{\fam\bmfam\twelvebm}
}

\def\tenpointbmit{
\textfont\bmfam=\tenbm
\scriptfont\bmfam=\seveni
\scriptscriptfont\bmfam=\fivei
\def\bmit{\fam\bmfam\tenbm}
}

\tenpointbmit

\mathchardef\Gamma="7100
\mathchardef\Delta="7101
\mathchardef\Theta="7102
\mathchardef\Lambda="7103
\mathchardef\Xi="7104
\mathchardef\Pi="7105
\mathchardef\Sigma="7106
\mathchardef\Upsilon="7107
\mathchardef\Phi="7108
\mathchardef\Psi="7109
\mathchardef\Omega="710A
\mathchardef\alpha="710B
\mathchardef\beta="710C
\mathchardef\gamma="710D
\mathchardef\delta="710E
\mathchardef\epsilon="710F
\mathchardef\zeta="7110
\mathchardef\eta="7111
\mathchardef\theta="7112
\mathchardef\iota="7113
\mathchardef\kappa="7114
\mathchardef\lambda="7115
\mathchardef\mu="7116
\mathchardef\nu="7117
\mathchardef\xi="7118
\mathchardef\pi="7119
\mathchardef\rho="711A
\mathchardef\sigma="711B
\mathchardef\tau="711C
\mathchardef\upsilon="711D
\mathchardef\phi="711E
\mathchardef\cho="711F
\mathchardef\psi="7120
\mathchardef\omega="7121
\mathchardef\varepsilon="7122
\mathchardef\vartheta="7123
\mathchardef\varpi="7124
\mathchardef\varrho="7125
\mathchardef\varsigma="7126
\mathchardef\varphi="7127



\vskip 2cm

\centerline{\bigbold BLACK HOLES WITH WEYL CHARGE}\vskip 0.7cm
\centerline{\bigbold AND}\vskip 0.7cm
\centerline{\bigbold NON-RIEMANNIAN WAVES}\vskip 0.7cm
\bigskip\bigskip\bigskip

\centerline{Robin W Tucker}
\medskip
\centerline{Charles Wang}
\medskip

\centerline{\it School of Physics and Chemistry,}
\centerline{\it University of Lancaster,
		Bailrigg, Lancs. LA1 4YB, UK}
\centerline{\tt r.tucker{\rm @}lancaster.ac.uk}
\centerline{\tt c.wang{\rm @}lancaster.ac.uk}

\vskip 1cm
\vskip 2cm

\bigskip\bigskip\bigskip\bigskip

\centerline{\bf ABSTRACT}
\vskip 1cm

\midinsert
\narrower\narrower\noindent


A simple modification to Einstein's theory of gravity
in terms of a non-Riemannian connection is examined. A new
tensor-variational   approach  yields  field
equations that possess a covariance similar to the gauge
covariance of electromagnetism.  These equations are shown to
possess solutions analogous to those found in the Einstein-Maxwell
system. In particular  one finds gravi-electric and gravi-magnetic
charges contributing to a spherically symmetric static
Reissner-Nordstr\"om metric. Such  Weyl ``charges'' provide a source
for the non-Riemannian torsion and metric gradient fields
instead of the electromagnetic field.  The theory suggests that matter
may be endowed with gravitational charges that couple to gravity in
a manner analogous to electromagnetic couplings in an electromagnetic
field. The nature of gravitational coupling to spinor matter in
this theory  is also  investigated and  a solution exhibiting
a plane-symmetric gravitational metric wave coupled via
non-Riemannian waves to a propagating spinor field is presented.

\endinsert




\vfill
\eject

\headline={\hss\rm -~\folio~- \hss}     

\def\frac#1#2{{#1\over #2}}

\Section{Introduction}

\twelvepointbmit

\def\TT{\ST_{[{\bf ric}]}{}  }

\def\td{\tilde}

\def\b{\beta}
\def\dd{{\hbox{d}}}
\def\DD{{\hbox{D}}}
\def\gma{{\gamma}}
\def\g{{\bmit\gamma}}
\def\Cal{\cal}
\def\wd{\wedge}

\def\R#1#2{R^#1{}_#2}

\def\Q#1#2{Q^#1{}_#2}

\def\Tor{{\bf T}}
\def\dotTor{{\bf {\dot T}}}

\def\SS{{\bf S}}
\def\dotSS{{\bf \dot S}}
\def\nab#1{\nabla_{{}_#1}}
\def\i#1{i_{{}_{#1}}}
\def\ST{{\cal T}}

\def\bfR#1#2{ {\bf R}_{{}_{#1,#2}}  }
\def\dotbfR#1#2{ {\bf \dot{R}}_{{}_{#1,#2}}  }
\def\bfg{{\bf g}}

\def\frac#1#2{{#1\over #2}}
\def\ot{\otimes}
\def\CC{{\cal C}}

In the absence of matter Einstein's theory of gravity allows an elegant
formulation in terms of (pseudo-)Riemannian geometry.
  The field equations follow as the local extremum of an action integral under
metric variations.  The integrand of this action is
simply  the curvature scalar associated with the  curvature of the
Levi-Civita connection times the (pseudo-~)Riemannian volume form of spacetime.
Such a connection $\nabla$
 is torsion-free and metric compatible.
Thus for all vector fields $X,Y$ on the spacetime manifold, the tensors
given by:
$$ \Tor(X,Y)=\nab{X}Y-\nab{Y}X-[X,Y]\Eqno $$
$$\SS=\nabla \bfg\Eqno $$\label\delg
are zero
where
 $ \bfg$ denotes the metric tensor, $\Tor$ the 2-1 torsion tensor
and ${\SS}$ the gradient tensor of $\bfg$ with respect to $\nabla$.
Such a Levi-Civita connection provides a useful reference
connection since it depends entirely on the metric structure of the
manifold. Einstein's description  provides  a well tested
theoretical edifice for describing the large scale structure of gravitational
phenomena and although numerous supergravity and superstring variants of
the theory have hinted at a more general geometry there has been little
evidence to suggest that the additional torsion and metric-gradient fields
can be given an immediate interpretation. Although Weyl
\Ref{ H Weyl, Geometrie und Physik,
  Naturwissenschaften {\bf 19} (1931) 49}\label\weylref
 saw the potential
inherent in certain non-Riemannian geometries his efforts to relate the
metric-gradient to the electromagnetic field were regarded as unsuccessful
and in the light of subsequent developments the unification of the
fundamental interactions has been sought elsewhere. In particular the
promise of string unification raises questions concerning the viability of
a classical geometrical description on all scales. At the level of
effective theories however (where one has a chance of testing their
predictions) there are hints that a non-Riemannian geometry may offer a
more economical and more elegant description of gravitational
interactions
\Ref{J Scherk, J H Schwarz, Phys. Letts {\bf 52B} (1974) 347},
\Ref{ T Dereli, R W Tucker, An Einstein-Hilbert Action for Axi-Dilaton
Gravity in \break 4-Dimensions, Lett. Class. Q. Grav. To Appear},
\Ref{ T Dereli, M \"Onder,  R W Tucker, Solutions for Neutral  Axi-Dilaton
Gravity in 4-Dimensions, Lett. Class. Q. Grav. To Appear},
\Ref{ T Dereli, R W Tucker,   Class. Q. Grav. {\bf 11} (1994) 2575}.

It appears that low energy dilaton and axi-dilaton interactions can be
accommodated in terms of a connection that gives rise to a
particular torsion and metric-gradient field. Whether one should treat
seriously such models in astrophysical contexts is open to debate but
  it is of interest to
enquire whether  non-Riemannian fields could give rise to observational
effects
\Ref{F W Hehl, J D McCrea, E W Mielke, Y Ne'eman: ``Metric-affine
gauge theory of gravity: field equations, Noether identities,
world
spinors, and breaking of dilation invariance''.
Physics Reports, To Appear (1995).
}\label\Hehl,
\Ref{ F W Hehl, E Lord, L L Smalley, Gen. Rel. Grav. {\bf 13} (1981) 1037},
\Ref{P Baekler,  F W Hehl, E W Mielke  ``Non-Metricity and Torsion'' in
Proc. of 4th Marcel Grossman Meeting on General Relativity, Part A,
Ed. R Ruffini  (North Holland 1986) 277},
\Ref{V N Ponomariev, Y Obukhov, Gen. Rel. Grav. {\bf 14} (1982) 309},
\Ref{ J D McCrea, Clas. Q. Grav. {\bf 9} (1992) 553},
\Ref{ A A Coley, Phys. Rev. {\bf D27} (1983) 728},
\Ref{ A A Coley, Phys. Rev. {\bf D28} (1983) 1829, 1844},
\Ref{ A A Coley, Nuov. Cim. {\bf 69B} (1982) 89},
\Ref{M Gasperini, Class. Quant. Grav. {\bf 5} (1988) 521},
\Ref{J Stelmach, Class. Quant. Grav. {\bf 8} (1991) 897},
\Ref{ A K Aringazin, A L Mikhailov, Class. Q. Grav. {\bf 8} (1991) 1685},
\Ref{ J-P Berthias, B Shahid-Saless, Class. Q. Grav. {\bf 10} (1993) 1039}.
In this paper we seek a {\it simple}
modification to the traditional Einstein-Hilbert action that provides a
dynamical prescription for a non-Riemannian geometry that might be considered
as a viable alternative to Einstein's metric theory. A minimal requirement
in this direction is the existence of a
   static spherically symmetric solution that can compete
with the Schwarzschild metric in such a  theory.

 There exists a large body of literature that approaches non-Riemannian
 theories of gravitation in the language of affine structure groups of
principal fibre bundles over spacetime. In this article we offer an
alternative
language in terms of the metric $\bfg$ and the connection $\nabla$.
 Such a connection is in 1-1 correspondence with the
notion of a $gl(4,R)-$ algebra valued connection on the bundle of linear
frames over spacetime. However, unlike the metric-affine gauge approach
\Hehl\ ,{\Ref{F W Hehl, J D McCrea, E W Mielke, Y Ne'eman, Found. Phys. {\bf
19}
(1989) 1075}
we
do not require  a reduction of the  $R^4 \times Gl(4,R)$ gauge group to
formulate the theory, relying rather on the traditional definitions above of
torsion and non-metricity in terms of $\nabla$ and the metric $\bfg$.
A notable feature of the underlying gauge symmetry of the actions discussed
in this paper
is its similarity to the gauge group of electromagnetism. Although of
gravitational origin, the metric tensor remains invariant just as in
Maxwell's $U(1)$ gauge covariant theory while $\nabla$ experiences a
transformation.
This distinguishes our gauge symmetry
from Weyl's original theory
\weylref\  in which the connection remained invariant under
a dilation of the gravitational metric and a transformation of the
non-metricity tensor.
The papers by Gregorash and Papini
\Ref{D Gregorash, G Papini, Il Nuo. Cim. {\bf 55A} (1980) 37},
\Ref{D Gregorash, G Papini, Phys. Letts. {\bf 82A} (1981) 67}
are concerned with extensions of Weyl's original theory as interpreted
by Dirac
\Ref{P A M Dirac, Proc. R. Soc. Lond. {\bf 333} (1973) 403}
and construct conformally invariant actions with the aid of additional
scalar fields that compensate the transformation of the curvature scalar
induced
by a local scaling of the  metric.
Many models in the metric-affine gauge approach are similarly based
on ``conformally invariant'' actions in which the metric also transforms
under the gauge group.
In the approach below the emphasis is on the analogy with the
 electromagnetic gauge group rather than  conformal invariance. Since
the metric remains inert under our gauge group we  designate the associated
conserved charges ``Weyl charges'' rather than ``dilation charges''.

In section 2 we summarise the essential prerequisites of non-Riemannian
geometry from a modern viewpoint. In section 3 we
describe a tensorial variational principle.
Working tensorially offers a number of advantages over traditional
variational techniques. No reliance is made on particular fields of frames
or coframes and the resulting formulae, once derived,
 are unambiguous and easy to apply.  Transition to more
traditional component oriented manipulations is of course possible but
great care must then be exercised in raising and lowering indices with the
metric since the latter  is no longer regarded as covariantly constant.
In section 4 we explicitly derive our field equations in the absence of
matter.
Spherically symmetric static solutions are derived in section 5 and their
properties under a Weyl ``gauge'' group made manifest.
We interpret these solutions as Reissner-Nordstr\"om black holes with
gravitational ``Weyl charge''. We justify this nomenclature in section 6
which deals with the conservation of this charge.
In section 7 the nature of gravitational coupling to spinor
matter in this theory is investigated and  a solution exhibiting a
plane-symmetric gravitational
metric wave coupled via a non-Riemannian geometry to a propagating spinor
field is presented.

\Section{Non-Riemannian Geometry}

A linear connection on a manifold provides a covariant way to differentiate
tensor fields. It provides a type preserving derivation on the algebra
of tensor fields that commutes with
contractions. Such a connection  will be denoted $\nabla$.
Given an arbitrary local basis
of vector fields $\{X_a\}$ the most general
 linear  connection is specified locally
by a set of $n^2$ 1-forms  $\Lambda^a{}_b$ where $n$ is the dimension of the
manifold:
$$\nab{{X_a}} \,X_b=\Lambda^c{}_b (X_a)\, X_c .\Eqno $$
Such a  connection can be fixed by specifying a (2, 0)
 symmetric metric tensor
$\bfg$, \break a (2-antisymmetric, 1) tensor $\Tor$ and a (3, 0) tensor
$\SS$, symmetric
in its last two arguments. If we require that $\Tor$ be the torsion of
$\nabla$ and $\SS$ be the gradient of $\bfg$ then it is straightforward to
determine the connection in terms of these tensors. Indeed since $\nabla$
is defined to commute with contractions and reduce to differentiation on
scalars it follows from the relation
$$X(\bfg(Y,Z))=\SS(X,Y,Z)+\bfg(\nab{X}Y,Z)+\bfg(Y,\nab{X}Z)\Eqno$$
that
$$2\bfg(Z,\nab{X}Y)=X(\bfg(Y,Z))+Y(\bfg(Z,X))-Z(\bfg(X,Y))-$$
$$\quad\quad \bfg(X,[Y,Z])-\bfg(Y,[X,Z])-\bfg(Z,[Y,X])-$$
$$\quad\quad\quad \bfg(X,\Tor(Y,Z))-\bfg(Y,\Tor(X,Z))-\bfg(Z,\Tor(Y,X))-$$
$$\quad\quad \SS(X,Y,Z)-\SS(Y,Z,X)+\SS(Z,X,Y)\Eqno $$\label\conn
where $X,Y,Z$ are any  vector fields.
The general curvature operator $\bfR{X}{Y}$ defined in terms of $\nabla$
by
$${\bf R}_{X,Y}Z=\nab{X}\nab{Y}Z-\nab{Y}\nab{X}Z-\nab{{[X,Y]}}Z\Eqno $$
is also a type-preserving tensor derivation on the algebra of tensor fields.
The general (3, 1) curvature tensor ${\bf R}$ of $\nabla$ is defined by
$${\bf R}(X,Y,Z,\beta)=\beta(\bfR{X}{Y} Z)\Eqno$$
where $\beta$ is an arbitrary 1-form. This tensor gives rise to a set of
local curvature 2-forms
$$R^a{}_b(X,Y)=\frac{1}{2}\,{\bf R}(X,Y,X_b,e^a)\Eqno$$
where $\{e^c\}$ is any local basis of 1-forms dual to $\{X_c\},\quad $
$e^a(X_b)=\delta^a{}_b$.
In terms of the connection forms
$$\R{a}{b}=\dd \Lambda^a{}_b+\Lambda^a{}_c\wd \Lambda^c{}_b .
\Eqno$$\label\Rforms
In a similar manner the torsion tensor gives rise to a set of local torsion
2-forms $T^a$:
$$T^a(X,Y)\equiv \frac{1}{2}\, e^a(\Tor(X,Y))\Eqno $$
which can be expressed in terms of the connection forms as
$$T^a=\dd e^a+\Lambda^a{}_b\wd e^b.\Eqno$$
Since the metric is symmetric the tensor $\SS$ can be used to define a set
of local non-metricity 1-forms $Q_{ab}$ symmetric in their indices:
$$Q_{ab}(Z)=\SS(Z,X_a,X_b).\Eqno $$
It will prove useful in the following to make use of the exterior covariant
derivative $\DD$. Its definition may be found in
\Ref{ I M Benn, R W Tucker, {\bf An Introduction to Spinors and Geometry
with
Applications in Physics}, (Adam Hilger) (1987)}\label\book.
With $g_{ab}\equiv \bfg(X_a,X_b)$,
it follows from
\delg\ that
$$Q_{ab}=\DD g_{ab}\Eqno $$
$$Q^{ab}=-\DD g^{ab}\Eqno $$
where, as usual, indices are raised and lowered with components of the metric
in the ambient local basis or cobasis.
We shall denote the metric trace of these forms
$$Q\equiv Q^a{}_a\Eqno $$\label\Weyl
and refer to this as the Weyl 1-form.
Riemannian geometry  chooses a metric-compatible torsion-free connection
where $\SS$ and $\Tor$ are zero.
A geometry with a torsion-free connection that preserves a conformal metric
is known as a Weyl space
\Ref{H Pedersen, K P Todd, Adv. in Mathematics {\bf 97} (1993) 74}.

It is sometimes useful to decompose $\Lambda^a{}_b$ into its Riemannian and
non-Riemannian parts, $\Omega^a{}_b$ and $\lambda^a{}_b$ respectively:
$$\Lambda^a{}_b=\Omega^a{}_b+\lambda^a{}_b.\Eqno$$
It follows from \conn\ that
$$2\Omega_{ab}=(g_{ac}i_b-g_{bc}i_a+e_ci_ai_b)\,\dd e^c+
(i_b\dd g_{ac}-i_a\dd g_{bc})\,e^c+\dd g_{ab}\Eqno$$\label\RConn
and
$$
2\lambda_{ab}=i_a\,T_b-i_b\,T_a-(i_ai_b T_c +
                i_bQ_{ac}-i_aQ_{bc})\,e^c-Q_{ab}\Eqno $$\label\NRConn
where, for any frame $\{X_a\}$, we abbreviate  the contraction operator
with respect to $X_a$ by $i_a$.

\Section{The Variational Principle}

\def\CC{{\cal C}}

We wish to find equations for the local extrema of action
functionals ${\cal S}$ of
a general connection and tensor fields.
We consider 1-parameter deformations of these fields in determining the
variational equations in the traditional manner and denote the variational
derivative of a tensor $W$  with respect to a field $F$
by $\underbrace{\dot W}_F$.
Since a specification of a metric $\bfg$ and a connection $\nabla$
determines a non-Riemannian geometry we shall first consider functionals
${\cal S}[\bfg,\nabla]$. If we  take
${\cal S}[\bfg,\nabla]=\int \Lambda(\bfg,\nabla)$ for
some $n$-form $\Lambda$   the field equations of the theory follow from
\def\und{\underbrace}
$$\und{\dot\Lambda}_\bfg=0\Eqno$$
$$\und{\dot\Lambda}_\nabla=0\Eqno$$
where the variations of $\Lambda$ are induced respectively from the tensor
variations
$${\bf h}=\dot\bfg\Eqno$$
in the metric and
$$ \g={\dot\nabla}\Eqno $$
in the connection.
Since we shall consider functionals constructed from the curvature tensor
and its various contractions we need a basic formula for the variation of
the curvature operator.
Regarding $\nabla$ as a curve in the space of all connections its tangent
induces a variation in the curvature operator given by:
$$\underbrace{{\dotbfR{X}{Y}Z}}_\nabla=\dot\nabla_X\,\nabla_Y\, Z-
\dot\nabla_Y\,\nabla_X\, Z +
\nabla_X\,\dot\nabla_Y\, Z- \nabla_Y\,\dot\nabla_X\, Z
-\dot\nabla_{[X,Y]}Z .\Eqno$$
Since $\dot\nabla$ is a (2, 1) tensor, which we have denoted by $\g$, then
$$\underbrace{{\dotbfR{X}{Y}Z}}_\nabla=({\nab{X}}\g)(Y,Z,-)-
        ({\nab{Y}}\g)(X,Z,-)+\g(\Tor(X,Y),Z,-)\Eqno $$
This formula simplifies considerably if we introduce a set of local 1-forms
$\gma^a{}_b$ by
$$\g(X,X_b,e^a)=\gma^a{}_b(X)\Eqno $$
and write it in terms of the exterior covariant derivative
$$e^a(\dotbfR{X}{Y} X_b)=2(\DD \gma^a{}_b)(X,Y).\Eqno $$
Finally we denote by $\star $ the Hodge map of the metric $\bfg$
and write the $n$-form measure $\star 1$ as
$\mu_\bfg$ to remind us of its metric
dependence.
It is a standard result that
$$\und{\dot\mu_\bfg}_\bfg=\frac{1}{2}\,{\bf h}(X_a,X^a)\mu_\bfg .\Eqno$$

\Section{The Field Equations}

Einstein's field equations in the absence of matter follow as the
variational equations deduced from the Einstein-Hilbert action. This is the
integral of the  curvature scalar
of the Levi-Civita connection. This scalar is obtained by
a metric contraction of the Ricci tensor. The latter is  a
trace of the curvature operator. When the metric  has a gradient and the
connection is not torsion-free there are basicly two distinct
second rank tensors that can be obtained by taking different traces:
$${\bf Ric}(X,Y)=e^a(\bfR{X_a}{X} Y)\Eqno $$
$${\bf ric}(X,Y)=e^a(\bfR{X}{Y} X_a) .\Eqno $$
The first has no symmetry in general
 while $\bf ric$ is a 2-form.
It follows from \RConn,\ \NRConn\ and \Rforms\ that
$${\bf ric}=2R^a{}_a=-\dd Q.\Eqno$$\label\ricdQ
The symmetric part
of $\bf Ric$ can be contracted with the symmetric metric tensor to define a
generalised curvature
$${\Cal R}={\bf Ric}(X_a,X_b){\bf g}(X^a,X^b).\Eqno $$
In view of the comments in the introduction concerning the gauge structure
of Maxwell's action for electromagnetism, by analogy
we adopt as  the gravitational action functional  the integral
$$\int\Lambda=
\int \kappa_1\,{\cal R} \mu_\bfg +\kappa_2 \, {\bf ric}\wd \star {\bf ric}
\Eqno $$\label\action
where $\kappa_1$ and $\kappa_2$ are couplings. Regarding this as a
functional of $\bfg$ and $\nabla$ we may exploit the variational
derivatives in the last section to deduce field equations for the metric
and connection. The relation
$$\underbrace{ \left( {{\bf ric}}
 \right) \dot{} \wd \star  {\bf ric}  }_{\nabla}=
4\,\DD \gamma^a{}_a\wd \star {\bf ric} $$
$$\quad = 4\,\gamma^a{}_a\wd \dd \star {\bf ric}
+ 4\,\dd (\gamma^a{}_a\wd \star {\bf ric})\Eqno$$
isolates the connection variation from the second term in the action. The
connection variation of the generalised Einstein-Hilbert term is
$$ \und{\dot{\cal R} \star 1}_\nabla=
g^{ab}\, \und{e^c(\dotbfR{X_c}{X_x}X_b)}_\nabla \star 1= g^{ab}
(\DD \gma^c{}_b)(X_c,X_a)\star 1=g^{ab}(\i{X_a}\i{X_c}\DD \gma^c{}_b) \star 1$$
$$=\quad -g^{ab} \DD \gma^c{}_b\wd \star (e_a\wd e_c)=
-\gma^c{}_b\wd \DD (\star (e^b\wd
e_c))-\dd (\gma^c{}_b\wd \star (e^b\wd e_c)).\Eqno $$
Hence for variations with compact support:
$$\int\underbrace{\dot\Lambda}_{\nabla}=\int
-\kappa_1\,\gma^b{}_a\wd \DD \star (e^a\wd
e_b)+4\kappa_2\,\gma^a{}_a \wd \dd \star {\bf ric} .\Eqno $$
In a similar manner  the metric variations of the generalised
Einstein-Hilbert  term give
$$\und{({\cal R} \star 1)\dot{}}_\bfg=
-({\bf Ric})(X^b,X^c)\,{\bf h}(X_b,X_c)\mu_\bfg
+ {\cal R}\und{\dot\mu_\bfg}_\bfg$$
$$\quad = -{\bf h}(X_b,X_c)\, {\bf Ein}(X^b, X^c) \mu_{\bfg}\Eqno$$
while the second term in the action  contributes a term involving
the ${\bf ric}$ stress tensor $\TT$:
$$\int\underbrace{\dot\Lambda}_{\bfg}=
  -\int {\bf h}(X^b,X^c)\,(\kappa_1\,{\bf Ein}(X_b,X_c)
     +\kappa_2\,\TT(X_b,X_c))\mu_\bfg .\Eqno $$
In these relations  $\hbox{ric}_{ab}\equiv {\bf ric}(X_a,X_b)$ and
$${\bf Ein}=\hbox{Sym}({\bf Ric})-\frac{1}{2}\, {\bfg} {\cal R}\Eqno $$
$${\tau_{[{\bf ric}]}}{}_a\equiv\,\frac{1}{2}\,
( \i{X_a}{\bf ric}\wd \star {\bf ric}
-\i{X_a}\star {\bf ric}\wd {\bf ric})\Eqno $$
$$\TT_{ab}\equiv\TT(X_a,X_b)=
{\star ^{-1}}(e_a\wd {\tau_{[{\bf ric}]}}{}_b)\Eqno $$
$$\quad = -4\,\hbox{ric}_{ac}\,\,\hbox{ric}^c{}_b -
             g_{ab}\,\hbox{ric}_{cd}\,\,\hbox{ric}^{cd}.\Eqno $$
Thus the field equations are
$$\kappa_1\,{\bf Ein}+\kappa_2\,\TT=0\Eqno $$\label\Eineq
$$\CC_b{}^a = 0 \Eqno $$\label\cart
where $$\CC_b{}^a =
-\kappa_1\,\DD\star (e^a\wd e_b)
+4\,\delta^a{}_b\,\kappa_2\,\dd \star {\bf ric}.\Eqno$$
Equation \cart\  gives immediately
$$\dd \star {\bf ric}=0\Eqno $$\label\Cartone
$$\DD \star (e^a\wd e_b)=0 .\Eqno $$\label\Carttwo
\def\FF{{\cal F}}
\def\i{i}
\def\Q{\widehat{Q}}
\def\T{\widehat{T}}
In general, the variational derivative of the integral
$$\int {\cal R} \mu_\bfg +{\FF}(\bfg,\nabla,\ldots)\Eqno$$
with respect to the
connection
gives rise to the equation
$$
\DD \star  (e^a \wedge e_b) = \FF^a_{\;\;b} \Eqno\label\Cartthree
$$
for some $(n-1)$-forms $\FF^a_{\;\;b}$ defined by
$$\underbrace{\dot\FF}_{\nabla}=\gma^b{}_a\wd \FF^a_{\;\;b}. \Eqno $$
Clearly this equation implies
$$
\FF^a_{\;\;a}=0. \Eqno
$$\label\trf
It is convenient to decompose $Q_{ab}$ into its trace and trace-free parts:
\Ref{J D McCrea, Class. Quant. Grav. {\bf 9}, (1992)  553}
$$
Q_{ab}=\widehat{Q}_{ab}+{1\over n}\,g_{ab}\,Q  \Eqno
$$
so that $\Q^a{}_a=0$.
Equation \Cartthree\  can be then decomposed as
$$
\i_b\,\Q_a^{\;\;c}-\delta^c_{\;\;b}\,\i_d \Q_a^{\;\;d}
+(\delta^c_{\;\;b}\,\delta^d_{\;\;a}
 -\delta^c_{\;\;a}\,\delta^d_{\;\;b})\,
({n-2\over 2n}\,\i_d Q-\i_{d}\,\i_{h}\,T^{h})
-\i_{b}\,\i_{a}\,T^{c}+f^c_{\;\;ab}=0
\Eqno
$$
that is,
$$
\i_a\,\Q_{bc}-\i_{a}\,\i_{b}\,T_{c}=
-{1\over 2n}\,g_{bc}\,\i_a\,Q+{1\over 2n}\,g_{ac}\,\i_b\,Q
-f_{cba}
\quad\quad\quad\quad\quad\quad\quad\quad
\quad\quad\quad\quad\quad\quad\quad\quad
$$
$$
\quad\quad\quad\quad
-{1\over n(n-2)}\,g_{ac}\,f^d_{\;\;db}
+{n-1\over n(n-2)}\,g_{bc}\,f^d_{\;\;da}
+{n-1\over n(n-2)}\,g_{ac}\,f^d_{\;\;bd}
-{1\over n(n-2)}\,g_{bc}\,f^d_{\;\;ad}
\Eqno
$$
where $\FF^a{}_b=f^{ca}{}_b \star e_c$.
Using the symmetry of $\Q_{bc}$ and the antisymmetry of
$\i_{a}\,\i_{b}\,T_{c}$ in $a\,b$,
and splitting off the trace-free part from the torsion form:
$$
T^a=\T^a+{1\over n-1}\,e^a\wedge T \Eqno
$$
where
$$
T\equiv \i_a\,T^a \Eqno
$$
such that $\i_a\,\T^a=0$,
it follows that
$$
\i_a\,\Q_{bc}
={1\over n}\,g_{bc}\,(f^d_{\;\;da}+f^d_{\;\;ad})
-{1\over2}\,(f_{bac}+f_{bca}+f_{cab}+f_{cba}-f_{abc}-f_{acb})
\Eqno
$$\label\Qhatsol
$$
\i_{a}\,\i_{b}\,\T_{c}
={1\over n-1}\,(g_{bc}\,f^d_{\;\;ad}-g_{ac}\,f^d_{\;\;bd})
-{1\over2}\,(f_{bac}+f_{bca}+f_{cab}-f_{cba}-f_{abc}-f_{acb})
\Eqno
$$\label\Torhatsol
$$
T-{ n-1\over 2n}\,Q
=\left ({1\over n(n-2)}\,f^c_{\;\;ac}- { n-1 \over n(n-2)}\,f^c_{\;\;ca}
\right )\,e^a
.\Eqno
$$\label\TQsol
The equations \Qhatsol\ , \Torhatsol\  and \TQsol\  provide the general
solution
to \Cartthree.\  It will be observed that the trace-free parts are
determined in terms of  contributions from $\FF$
 while the trace of the torsion
and Weyl forms remain correlated. If the contributions
from $\FF$ are algebraic in the components of the torsion and metric-gradient
these equations enable one
to immediately express the connection in terms of the
metric, the components of $\star \FF^a{}_b$ and the Weyl form $Q$.

Before analysing the field equations \Cartone\  and  \Carttwo\
corresponding to
 $\FF=\kappa_2/\kappa_1\,{\bf ric} \wd \star  {\bf ric}$,
it is of interest to verify  that
\Cartthree\ is equivalent to the variational equations of
any action regarded as the  functional of $\bfg ,\, \Tor$ and $\SS$ obtained
by replacing the connection explicitly by \conn.\  We note that the
 $(n(n+1)+2n^3)/2$ component
variations of the metric and connection may be considered as induced by the
$n(n+1)/2$
component metric variations, the
$n^2(n+1)/2$  component variations of the metric-gradient tensor and the
$n^2(n-1)/2$
component variations of the torsion tensor in
$n$ dimensions.
Thus the $\gma^a{}_b$ variations   in
$$\int\und{\dot\Lambda}_{\bfg,\Tor,\SS}=
  -\int {\bf h}(X^b,X^c)\,(\kappa_1\,{\bf Ein}(X_b,X_c)
     +\kappa_2\,\TT(X_b,X_c))\mu_\bfg
+\gma^a{}_b\wd \CC_a{}^b. \Eqno $$
are to be induced by variations ${\bf h}, \, \dot\Tor$ and $\dot\SS$ of
${\bfg}, \, \Tor$ and $\SS$ respectively.
By varying \conn\ we readily find
$$\g(X,Y,\b)=\underbrace{\g(X,Y,\b)}_{\Tor} +
\underbrace{\g(X,Y,\b)}_\SS +
\underbrace{\g(X,Y,\b)}_\bfg \Eqno$$
where
$$2\underbrace{\g(X,Y,\b)}_{\Tor}=
g(X,\dotTor(\tilde\b,Y)+g(Y,\dotTor(\tilde\b,X)+
              g(\tilde\b,\dotTor(X,Y))\Eqno $$
$$2\underbrace{\g(X,Y,\b)}_\SS=\dotSS(\td\b,X,Y)-
\dotSS(X,Y,\td\b)-\dotSS(Y,\td\b,X)\Eqno $$
$$2\underbrace{\g(X,Y,\b)}_\bfg=(\nab{X} {\bf h})(Y,\td\b)
+(\nab{Y} {\bf h})(X,\td\b)   -(\nab{{\td\b}} {\bf h} )(X,Y)\Eqno $$
       $$\quad\quad\quad -{\bf h}(X,\Tor(Y,\td\b))-{\bf h}(Y,\Tor(X,\td\b))
                  -{\bf h}(\td\b,\Tor(Y,X))\Eqno $$
for any vector fields $X,Y$ and 1-form $\beta$ with metric dual $\td\b$
defined by $g(\td\b,Y)=\b(Y)$.
Writing
$$\gma^a{}_b\wd  \CC_a{}^b=\g(X_c,X_b,e^a)\,e^c\wd \CC_a{}^b$$
we may express these relations in terms of $h_{ab}={\bf h}(X_a,X_b)$,
$T_{ab}{}^c=e^c(\Tor(X_a,X_b))$,
$S_{abc}=\SS(X_a,X_b,X_c)$
and
$\gma^a{}_b(X_c)= \gma_{cb}{}^a$ where
$$\gma_{cba}
=\frac{1}{2}\,\{\dot T_{abc}+\dot T_{acb}+\dot T_{cba}\}+
\frac{1}{2}\,\{-\dot S_{cba}-\dot S_{bac}+\dot S_{acb}\}+$$
$$
\frac{1}{2}\,\{(\DD h_{ab})(X_c)+(\DD h_{ac})(X_b)-(\DD h_{bc})(X_a)
-h_{ck}\, T_{ba}{}^k-h_{bk}\, T_{ca}{}^k-h_{ak}\, T_{bc}{}^k\}. \Eqno$$
Thus in terms of the $n$-forms
$\CC_a{}^{bc}\equiv e^c\wd \CC_a{}^b$ in $n$-dimensions:
$$\gma^a{}_b\wd \CC_a{}^b=\gma_{cb}{}^a\CC_a{}^{bc}$$
$$=\frac{1}{2}\,\dot T_{abc}\{\CC^{abc}+\CC^{acb}+\CC^{cba}\}
+\frac{1}{2}\,\dot S_{abc}\{-\CC^{cba}-\CC^{bac}+\CC^{abc}\}+$$
$$
\frac{1}{2}\,h_{ab}\{ T_{ck}{}^b\, \CC^{cka}-T_{ck}{}^b\,
\CC^{kac}+T_{ck}{}^b\, \CC^{akc}\}+$$
$$+\frac{1}{2}\,(\DD h_{ab})(X_c)\{ \CC^{abc}+\CC^{acb}-\CC^{cab}\}.\Eqno$$
The independent torsion variations and metric-gradient variations give
respectively:
 $$\CC^{abc}+\CC^{acb}+\CC^{cba}-\CC^{bac}-\CC^{bca}-\CC^{cab}=0\Eqno$$
$$\CC^{abc}+\CC^{acb}-\CC^{cba}-\CC^{bac}-\CC^{bca}-\CC^{cab}=0.\Eqno$$
Hence we immediately deduce that
$\CC^{abc}=0$
corresponding to
$$\CC_b{}^a=0\Eqno $$
derived above.
Furthermore the metric variations now reproduce
the same Einstein equation so
one recovers the same
field equations for the metric, metric-gradient and torsion.


\Section{General Solutions}

We first find the general solution to \Carttwo\ . This is an algebraic
equation relating the components of the metric-gradient and the torsion
tensor. It has less than maximal rank and from \trf\ and  \TQsol\
with the
action \action\  it follows that the general solution is
expressible in terms of an arbitrary  Weyl form $Q=\alpha$. In $n$-dimensions:
$$Q_{ab}=\frac{1}{n}\, g_{ab}\, \alpha\Eqno $$\label\Qsol
$$T^a=\frac{1}{2n}\, e^a\wd \alpha\Eqno $$\label\Torsol
corresponding to the
tensors
$$\SS=\frac{1}{n}\, \alpha \otimes \bfg\Eqno $$
$$\Tor=-\frac{1}{n}\,(\alpha \wd e^a)\otimes X_a.\Eqno $$
The choice $\alpha=0$ corresponds to (pseudo-)Riemannian geometry
 with ${\bf ric}=0$ and reduces the theory to Einsteinian gravity for $n$=4.
The remaining system \Eineq\ and \Cartone\ is reminiscent of the
Maxwell-Einstein system.
Indeed it is not difficult to show that for any
torsion and metric-gradient our action \action\ is invariant under the
``gauge'' transformations:
$$ \bfg \mapsto \bfg$$
$$\nabla\mapsto\nabla+\dd f\otimes\Eqno$$\label\gauge
when $\nabla$ acts on vector fields and  $f$ is any 0-form.
This implies that the curvature operator and hence ${\cal R}$ and ${\bf
ric}$
are gauge invariant.
In a fixed local coframe the connection forms of $\nabla$:
$$\Lambda^c{}_b\mapsto \Lambda^c{}_b+ \delta^c{}_b \, \dd f.\Eqno$$
Furthermore the Bianchi identity $\dd\, {\bf ric}=0$ implies that locally
${\bf ric}=-\dd \alpha$ for some 1-form $\alpha$.
{}From \ricdQ\  we may
identify the  potential $\alpha$  with the  Weyl form
$Q$. It is important to stress that in contradistinction to
electromagnetism where the gauge potential is identified with a 1-form in
the Lie algebra of the compact $U(1)$ group, the 1-form $\alpha$ is
determined by the dynamical non-Riemannian geometry associated with a
connection  in the Lie-algebra of the general linear group, the structure
group of the bundle of linear frames.
With the solutions \Qsol\
and \Torsol\ it follows that
$$\lambda_{ab}=-\frac{1}{8} g_{ab} Q\Eqno $$
and the tensor ${\bf Ein}$ reduces to the Levi-Civita Einstein tensor of
Einstein's theory. Consequently \Eineq\ is isomorphic to the standard
Einstein-Maxwell system with the gravitational field $d\, Q$ corresponding
to the Maxwell field strength. It follows that \Eineq\ admits {\it all}
solutions to that system with an appropriate correspondence between couplings.
For example, with $Q=\alpha$ we find the exact spherically symmetric static
 spacetime ($n=4$) solution to the remaining field equations
\Eineq\ and \Cartone\ :
$$\alpha=\frac{q\,\dd t}{r}\Eqno$$


$$ \bfg=-\left(1-\frac{2M}{r}-\frac{\kappa_2\,q^2}{2\kappa_1\, r^2}\right)
\, \dd t\ot \dd t +
      \left(1-\frac{2M}{r}-\frac{\kappa_2\, q^2}{2\kappa_1\,
r^2}\right)^{-1}\, \dd r\ot \dd r +$$
$$\quad\quad       r^2\, \dd \theta\ot \dd \theta
      + r^2 \,\sin^2\theta \dd \phi\ot \dd \phi \Eqno $$\label\metricsol
in a chart with coordinates $\{t,r,\theta,\phi\}$.
$M$ and $q$ are arbitrary constants and the metric is asymptotically flat.
For $q\neq 0$
our metric takes the form of the Reissner-Nordstr\"om solution where
the associated stress tensor is provided by contributions from the
Weyl form \Weyl.\
This particular solution has also featured in \Ref{R Tresguerres,
Z. f\"ur Physik {\bf C} 65 (1995) 347 }\label\added\  where a considerably
more complex non-Riemannian action has been analysed. We stress that the
vacuum field equations from the
action \action\ above admit solutions that may be
constructed from {\it all}  Einstein-Maxwell solutions.

If we define a geodesic  test particle to be one that follows a time-like
geodesic associated with the metric $\bfg$ then  both electrically
neutral and electrically charged  geodesic test particles would behave in
the same way in the geometry of this solution. While it is by no means
obvious that real test particles behave like geodesic test particles
this  phenomena suggests that massive particles may be endowed with a
``gravitational charge''  that couples to gravity in a manner similar to the
way electrical charge couples to the electromagnetic field.

\Section{Conservation of Weyl Charge}

\def\del{\nabla}
\def\inv{^{-1}}
\def\FFF{{\cal F}_{\vartheta,\varphi}}
\def\calD{{\cal D}}

We have identified the constant $q$ in the Reissner-Nordstr\"om solution
above with a new kind of ``Weyl charge''
 associated with the gauge transformation \gauge\
 above. This is by analogy with the identification of electric charge of a
black hole in the Einstein-Maxwell theory.
It is of interest to consider how the conservation of ``Weyl charge''
arises in the context of a model of Weyl charged scalar fields. To this end
 consider the covariant derivatives of the exterior product of $n$
orthonormal 1-forms $\star 1$ and $n$ orthonormal vector fields $\sharp 1$:
$$
\del \star 1 = {1\over2}\, Q \otimes \star 1
\Eqno
$$\label\delstarone
$$
\del \,\sharp\, 1 = -{1\over2}\, Q \otimes \,\sharp\, 1
\Eqno
$$\label\delsharpone
in terms of the Weyl form $Q$.
Any real
$n$-form $\Phi$ can be expressed in terms of the pseudo-scalar $\varphi$
with respect to $\star 1$ by:
$$
\Phi=\varphi \star 1.
\Eqno
$$\label\defPhi
Similarly any real $n$-multi-vector $\Theta$ can be expressed in
terms of the pseudo-scalar $\vartheta$
with respect to $\sharp1$ by:
$$
\Theta=\vartheta \,\sharp\, 1.
\Eqno
$$\label\defTheta
Using \delstarone\ and \delsharpone\ it follows that
$$
\del\Phi=\calD\varphi\otimes  \star 1
\Eqno
$$\label\delPhi
$$
\del\Theta=\bar\calD\vartheta\otimes  \sharp 1
\Eqno
$$\label\delTheta
where
$$
\calD \equiv \dd\, + \,{1\over2}\,Q\wd
\Eqno
$$\label\delcalD
$$
\bar{\calD} \equiv \dd\, - \,{1\over2}\,Q\wd
\Eqno
$$\label\delbarcalD
denote the appropriate  Weyl exterior covariant derivatives.
The action $n$-form:
$$
\FFF(\bfg,\nabla,\vartheta, \varphi)\,\equiv\,
\del_{X_a}\Phi\;\sharp\,\inv\,\del_{X^a}\Theta
\,+\, m^2\,\Phi\;\sharp\,\inv\,\Theta
\;=\; \calD \varphi \,\wd\star\, \bar{\calD}\vartheta
+ m^2\varphi\,\vartheta\star1
\Eqno
$$\label\lag
is invariant under the
 transformations:
$$
\bfg\mapsto \bfg\,,\;\;\;
\Phi\mapsto\exp(nf)\,\Phi\,,\;\;\;
\Theta\mapsto\exp(-nf)\,\Theta
\Eqno
$$\label\gaugelag
together with
$$\del\mapsto\del\,+\,\dd f\otimes $$
acting on vector fields.
These induce the corresponding transformations
$$
\varphi\mapsto\exp(nf)\,\varphi\,,\;\;\;
\vartheta\mapsto\exp(-nf)\,\vartheta \,,\;\;\;
Q\mapsto Q \,-\,2\,n\,\dd f.
$$
The field equations for $\varphi$ and $\vartheta$ follow from the variations
$$
\underbrace{\dot{\FFF}}_{\varphi}
=-\dot{\varphi}\,(\,\bar{\calD}\star\bar{\calD}\vartheta -
m^2\vartheta\star1\,)
\;+\;\dd(\dot{\varphi}\star\bar{\calD}\vartheta)
$$
$$
\underbrace{\dot{\FFF}}_{\vartheta}
=-\dot{\vartheta}\,(\,{\calD}\star{\calD}\varphi - m^2\varphi\star1\,)
\;+\;\dd(\dot{\vartheta}\star{\calD}\varphi)
$$
as
$$
\bar{\calD}\star\bar{\calD}\vartheta - m^2\vartheta\star1=0
\Eqno
$$\label\feqntheta
$$
{\calD}\star{\calD}\varphi - m^2\varphi\star1=0\,.
\Eqno
$$\label\feqnphi
In these equations we may identify the real constant $m$ as the mass of the
real ``Weyl doublet'' $(\varphi,\vartheta)$.

The metric and connection variations follow from the relations
$$
\underbrace{\dot{Q}}_{\bfg} = \dd h
\Eqno
$$\label\dotQg
$$
\underbrace{\dot{Q}}_{\del} = -2\gamma
\Eqno
$$\label\dotQdel
where
$$
h\equiv h^a{}_a
\Eqno
$$\label\defh
$$
\gamma\equiv\gamma^a{}_a.
\Eqno
$$\label\defgamma
If we now add the integral of the action density \lag\  to \action\
 and consider variations in $\nabla$
we generate the field equations
$$
4\kappa_2\,\dd\star{\bf ric}+{\bf j}=0
\Eqno
$$\label\Carone
$$
\DD\star(e^a\wd e_b)=0
\Eqno
$$\label\Cartwo
where
$$
{\bf j}\equiv\vartheta\star{\calD}\varphi\,
- \,\varphi\star\bar{\calD}\vartheta
\Eqno
$$\label\defj
follows from the variation
$$
\underbrace{\dot{\FFF}}_{\del}
=
{1\over2}\,\underbrace{\dot{Q}}_{\del}\,\varphi\wd\star\bar{\calD}\vartheta
\;-\;{1\over2}\,\underbrace{\dot{Q}}_{\del}\,\vartheta\wd\star{\calD}\varphi
=
-\gamma\wd(\,
\varphi\star\bar{\calD}\vartheta\,-\,\vartheta\star{\calD}\varphi\,).
$$
The equation \Cartwo\ may be solved as before to fix the connection.
The closure of ${\bf j}$ is compatible with
\feqntheta\ and \feqnphi\ since
$$
\dd\, {\bf j}
=
\dd\vartheta\wd\star{\calD}\varphi \;+\;
\vartheta\;\dd\star{\calD}\phi
\;-\;\dd\varphi\wd\star\bar{\calD}\vartheta
\;-\;\varphi\;\dd\star\bar{\calD}\vartheta
$$
$$
\!\!\!\!
=
\dd\vartheta\wd\star{\calD}\phi \;+\;
\vartheta\;(-{1\over2}\,Q\wd\star\calD \varphi + m^2\varphi\star1)
$$
$$
\;\;\;\;\;\;\;
-\,\dd\varphi\wd\star\bar{\calD}\vartheta
\;-\;\varphi\;({1\over2}\,Q\wd\star\bar{\calD} \vartheta + m^2\vartheta\star1)
$$
$$
=
\bar{\calD} \vartheta \wd\star{\calD}\varphi \,-\,
{\calD}\varphi\wd\star\bar{\calD} \vartheta
=
0.
$$
The  set of field equations is completed by taking  the metric variation:
$$
\!\!\!\!\!\!\!\!\!\!\!\!\!\!\!
\underbrace{\dot{\FFF}}_{\bfg}
=
\calD\varphi\wd\underbrace{\dot{\star}}_{\bfg} \bar{\calD}\vartheta
\;+\;m^2\,\varphi\,\vartheta\underbrace{\dot{\star 1}}_{\bfg}
\;+\;\underbrace{\dot{\calD}}_{\bfg}\varphi\wd\star\bar{\calD}\vartheta
\;+\;\calD\varphi\wd\star\underbrace{\dot{\bar{\calD}}}_{\bfg}\vartheta.
$$
$$
\;\;\;\;\;\;\;\;
=
-h^{ab}\,(\,i_{X_a}\,\calD\phi\wd\star i_{X_b}\bar{\calD}\theta
\,-\,g_{ab}\,\FFF\,)
\,+\,{1\over2}\underbrace{\dot{Q}}_{\bfg}\wd
(\,\phi\star\bar{\calD}\theta\,-\,\theta\star\calD\phi\,).
\Eqno
$$\label\varg
The term in $\dot Q$ does not contribute since ${\bf j}$ is closed:
$$
\underbrace{\dot{Q}}_{\bfg}\wd
(\,\varphi\star\bar{\calD}\vartheta\,-\,\vartheta\star\calD\varphi\,)
=
-\dd h \wd {\bf j}
=
h\,\dd {\bf j} - \dd(h\, {\bf j}).
$$
Thus the Einstein equation for $\bfg$ becomes
$$
\kappa_1{\bf Ein}+\kappa_2{\TT}=
{\cal T}\!_{[\vartheta,\varphi]}
$$
where
$$
\!\!\!\!\!\!\!\!\!
{\cal T}\!_{[\vartheta,\varphi]}
\equiv
-{1\over2}\left(
\calD\varphi\otimes\bar{\calD}\vartheta
+\bar{\calD}\vartheta\otimes\calD\varphi
-{\bfg}(\widetilde{\calD\phi},\widetilde{\bar{\calD}\theta})\,\bfg
-m^2\,\bfg \,\right)
\Eqno
$$\label\stresstf
and ${\bf Ein}$ remains the Einstein tensor associated with the Levi-Civita
connection.
Finally since ${\bf ric}=-\dd Q$:
$$
\dd\,{\bf ric}=0.
\Eqno
$$\label\dric
Thus we have demonstrated how the conservation of the ``Weyl current''
${\bf j}$ follows from a local
Weyl gauge covariant coupling of ``Weyl charged'' matter
fields to gravity.
This approach may be compared with the one given in
\Ref{R Hecht, F W Hehl, J D McCrea, E Mielke, Y Ne'eman, Phys. Letts. {\bf
A172} (1992) 13}
 which deals with  conservation laws induced by vector fields that
generate  {\it Killing
symmetries} in non-Riemannian spacetimes.

\Section{Gravitational Wave Coupling to Spinor Matter}

It is also of interest to examine the coupling of matter with zero Weyl
charge to non-Riemannian
gravity. To this end we examine the modification to the field equations
$\CC_b{}^a=0 $ produced when we consider a total action $n$-form
$$
\Lambda_{\hbox{Total}}=\Lambda+\FF_\Psi(\bfg,\nabla,\Psi) \Eqno
$$
\def\S{{\sl S}\/}
\def\Sslash{{{\S \!\!\! \slash}}}
As an example, consider a complex spinor field $\Psi$ in
spacetime with an action $n$-form
$$
\FF_\Psi=(\Psi\,,\Sslash\,\Psi) \mu_{\bfg}
.\Eqno
$$\label\spinaction
We shall employ the language of Clifford bundles to describe spinor fields
\book\
\Ref{I M Benn, A al-Saad, R W Tucker, Gen. Rel. Grav. {\bf 19} (1987)
1115},
although the transcription to the representation in terms of component
spinors in a $\gamma$-matrix language is straightforward
\Ref{J Schray, R W Tucker, C Wang, LUCY: A Clifford Algebra Approach to
Spinor Calculus, Lancaster University Preprint (1995)}.
The  {\it symmetric} inner product on spinors is defined as
\def\PP#1{{\cal P}_#1}
$$
(\Phi\,,\Psi)\equiv 4\,\hbox{Re}\, \PP0(\widetilde\Phi\vee\Psi)
\Eqno
$$
where $\PP{l}$ denotes the projection of a Clifford form to its
$l$-form component and $\vee$ denotes the Clifford product between sections of
the Clifford bundle associated with the spacetime metric.
The  familiar discussion of spinors is
effected in terms of a local $\bfg$
-{\it orthonormal} coframe $\{e^a\}$.
We choose as spinor adjoint
$
\widetilde\Psi\equiv C^{-1}\vee\Psi^J
$
with
$
C=\i\,e^0
$
and the adjoint involution
$
J=\xi\eta^*
$
where $\eta$ and $\xi$ are the main involution and anti-involution
of the Clifford algebra respectively and $*$ signifies the
complex conjugation.
The operator
$\Sslash\equiv e^a \vee {\S}_{X_a}$
is the Dirac operator which is expressed in
terms of a spinor covariant derivative ${\S}_X$.
Such a derivative may be expressed in
terms of a connection on the bundle of spinor frames, whose structure group
is a double cover of the spacetime Lorentz group.
In terms of the local $\bfg$
-{\it orthonormal} coframe it is expressible in terms of the
antisymmetric parts
of the connection forms $\{\Lambda_{ab}\}$.
For the action \spinaction\ one finds

$$
\und{\dot\FF_\Psi}_{\nabla}=\gma^a{}_b \wd
{1\over4}\,(\Psi\,,
[e^c\wd e_a\wd e^b]
\vee\Psi)\star   e_c.
\Eqno
$$

It follows from \Qhatsol,\  \Torhatsol\  and \TQsol\  that
 the traces of the torsion and non-metricity forms are
correlated  according
$$
T-{3\over8}\,Q=0
\Eqno
$$\label\newTQ
while the trace-free parts are specified to be
$$
\Q_{ab}=0
\Eqno
$$\label\newQh
$$
\i_{a}\,\i_{b}\,\T_{c}=
-{1\over 4 \kappa_1}\,(\Psi\,,[e_c\wd e_b\wd e_a]\vee\Psi)
\Eqno
$$
corresponding to the torsion forms:
$$
\T^{c}={1\over  \kappa_1}\hbox{Re}\,
\PP2(e^c\vee \Psi\vee \widetilde\Psi)
.\Eqno
$$\label\newTh

With $Q=\alpha$ for some 1-form $\alpha$
we must solve
$\FF^a{}_a=0$ or
$$\dd \star {\bf ric}=-\dd\star\,\dd \alpha=0\Eqno $$\label\riceqn
together with the  field equation
arising from the vanishing of the variation of  \spinaction\ induced by
 $\dot\Psi$:
 $$\und{\dot\FF_{\Psi}}_\Psi
=2\,(\dot\Psi,\Sslash\Psi)-\left(\frac{1}{2}\,i^c\Q_{ca}+(T-\frac{n-1}{2n}\,
Q)(X_a)\right)(\dot\Psi,e^a\vee \Psi)\mu_\bfg -
\dd\left((\dot\Psi,e^a\vee\Psi)\star e_a\right).
\Eqno$$

Thus from \newTQ\ and  \newQh\ the spinor field equation in spacetime
 becomes simply:
$$\Sslash\Psi=0.
\Eqno
$$\label\spineqn
{}From the metric variations one finds the Einstein equation:
$$\kappa_1\,{\bf Ein}+\kappa_2\,\TT=\ST_{[\Psi]}\Eqno $$\label\eeineqn
where
$$
\und{\dot\FF_\Psi}_{\bfg}= -{\bf h}(X^a,X^b)\ST_{[\Psi]}(X_a,X_b)
\mu_\bfg
\Eqno
$$
with
$$
\ST_{[\Psi]}\equiv
-{1\over 4}\,
(\Psi,[
 e_a \vee \S_{X_b}
+e_b \vee \S_{X_a}]\,\Psi)\,
e^a \otimes  e^b
-{1\over 2}\, (\Psi,\Sslash\Psi)\, \bfg.
\Eqno
$$
We seek a solution in a local coordinate system $\{u,v,x,y\}$ in which the
the metric takes the form
$$
g=H\,\dd u \otimes \dd u+\dd u \otimes \dd v+\dd v \otimes \dd u
+\dd x \otimes \dd x+\dd y \otimes \dd y.
\Eqno$$\label\geqn
In terms of a local orthonormal coframe
$$
g=-e^0 \otimes e^0+e^1 \otimes e^1+e^2 \otimes e^2+e^3 \otimes e^3
\Eqno$$\label\geqn
where
$$
e^0={H-1\over2}\,\dd u+\dd v
$$
$$
e^1={H+1\over2}\,\dd u+\dd v
$$
$$
e^2=\dd x
$$
$$
e^3=\dd y
\Eqno$$\label\cof
for some real function $H=H(u,x,y)$.
Such a metric describes a propagating plane-symmetric gravitational wave.
We seek a corresponding propagating tensor {\bf ric} with Weyl form
$$
\alpha=\rho\,\dd x+\zeta\,\dd y\Eqno
$$
where $\rho=\rho (u)$ and  $\zeta=\zeta (u)$ are real.
We find that, with the metric \geqn,\ the field equation \riceqn\ is satisfied
for arbitrary $\rho$ and $\zeta$.
We choose a complex spinor basis in a minimal left ideal of the complex
Clifford algebra of spacetime in which the element
$$
P={1\over4}\,(1+e^0 \vee e^1) \vee (1+i\,e^2 \vee e^3)
\Eqno$$
is a primitive idempotent. A spinor basis in this ideal can be chosen in
which a general spinor takes the form
$$
\Psi=(\Psi_1+\Psi_2\,e^2-\Psi_3\,e^0+\Psi_4\,e^0\vee e^2)\vee P
\Eqno
$$
where $
\Psi_1,\,\Psi_2,\,\Psi_3,\,\Psi_4
$
are complex functions and in which
the elements of the orthonormal coframe \cof\ are represented by the
$\gamma$-matrices:
$$
\gamma^0=
\pmatrix{ 0 & 0 & 1 & 0 \cr 0 & 0 & 0 & -1 \cr -1 & 0 & 0 & 0 \cr 0
 & 1 & 0 & 0 \cr}
$$

$$
\gamma^1=
\pmatrix{ 0 & 0 & 1 & 0 \cr 0 & 0 & 0 & -1 \cr 1 & 0 & 0 & 0 \cr 0
 & -1 & 0 & 0 \cr}
$$

$$
\gamma^2=
\pmatrix{ 0 & 1 & 0 & 0 \cr 1 & 0 & 0 & 0 \cr 0 & 0 & 0 & 1 \cr 0
 & 0 & 1 & 0 \cr}
$$

$$
\gamma^3=
\pmatrix{ 0 & i & 0 & 0 \cr {-}i & 0 & 0 & 0 \cr 0
 & 0 & 0 & i \cr 0 & 0 & {-}i & 0 \cr} .
\Eqno$$
In this basis, $\Psi$ becomes
$$
\Psi=\pmatrix{ \Psi_1 \cr \Psi_2 \cr \Psi_3 \cr \Psi_4 \cr}.
\Eqno
$$
{}From \newTQ\ and  \newQh\ with $Q=\alpha$ we may construct the
spinor covariant derivative that appears in \spineqn.\
This is a non-linear equation for the spinor components $\Psi_j$
 since $\Psi$
appears quadratically in the spinor connection.
However the propagating ansatz
$$
\Psi_1=0,\,\Psi_2=0
,\,\Psi_3=\Psi_3(u)
,\,\Psi_4=\Psi_4(u)
\Eqno
$$\label\ansatz
is found to be a particular solution for arbitrary
$\Psi_3(u)$ and $\Psi_4(u)$.
Finally the Einstein equation \eeineqn\
is satisfied if
$$
 {\kappa_1\over2}\,\left({{\frac {\partial ^{2}}{\partial x^{2}}}\
}+{{\frac {\partial ^{2}}{\partial y^{2}}}\,
}\right)H=
{\kappa_2}\,\left({\left({\frac {\partial\,\rho}
{\partial\,u}}\right)^2}
+{
\left({\frac {\partial\,\zeta}{\partial\,u}}\right)^2}\right)
+\frac{1}{2}\,{\hbox{Im}}
\left(\frac{\partial\Psi_3}{\partial\,u}\,\Psi_3^*+
      \frac{\partial\Psi_4}{\partial\,u}\,\Psi_4^*\right)\Eqno
$$\label\newHeqn

The metric function $H$ is determined by
this equation in terms of propagating spinor,  torsion and metric-gradient
waves with arbitrary profiles.

\Section{Conclusions}

The consequences of a generalisation of Einstein's metric
theory of gravitation has been examined
in terms of an action functional  dependent on a
general linear connection in additional to the spacetime metric.
A simple
generalisation of the variational principle with the Einstein-Hilbert
action  permits one to determine such a connection.
We have discussed
spherically symmetric static  solutions in which the Weyl form may be
interpreted in terms of a gravitational analogue of the Maxwell potential
and the Reissner-Nordstr\"om  metric
 arises in terms of a gravi-electric source.
Moreover we have shown that the vacuum theory admits solutions for the metric
corresponding to {\it all} those in
the standard Einstein-Maxwell theory.
For example a more general solution in the presence of the torsion
\Torsol\ and metric-gradient \Qsol\  can be generated from the Weyl form
$$\alpha= \frac{q\,\dd t}{r}+\mu \cos\theta \,\dd \phi\Eqno$$\label\gpole
where $\mu$ is a constant. The solution for the metric is then
 \metricsol\ with $q^2\mapsto q^2+\mu^2$.
The theory admits both
gravi-electric and gravi-magnetic poles with a duality symmetry analogous
to Einstein-Maxwell theory.

An interpretation of these solutions can be based on
 the singularity structure
of frame and gauge invariant tensors.
The tensor $T_a\wedge\star T^a$ is not gauge invariant.
However, since $\DD T^a=R^a{}_b\wd e^b$ and the curvature forms are gauge
invariant
under \gauge\ ,then $\DD T^a$ is also gauge invariant.
A gauge and  frame independent  invariant is
$$\DD T_a\wd \star \DD T^a=\frac{q^2-\mu^2}{32 \,r^4}.\Eqno$$
Similarly a gauge and  frame independent curvature invariant is
$$\star (R_{ab} \wd \star R^{ab})=
-\frac{24\,M^2}{r^6}+\frac{q^2-\mu^2}{16\,r^4}
-\frac{24\,\kappa_2 M\,(q^2+\mu^2)}{\kappa_1\, r^7}-
\frac{7\,\kappa_2^2\,(q^2+\mu^2)^2}{\kappa_1^2\,r^8}.\Eqno$$
The generic black hole
singularity at $r=0$ is clearly visible in these expressions.

The coupling of both Weyl charged   and Weyl
 neutral matter to this
generalised theory of gravitation  has also  been briefly examined.
The inclusion of Weyl neutral
spinor matter is of interest since the spin invariant
matter action is sensitive to the connection variation used to determine
dynamically  the
non-Riemannian geometry.
A family of solutions has been presented describing
 propagating Weyl spinor fields coupled to propagating metric,
torsion and metric-gradient plane symmetric waves.
The inclusion of the Maxwell action is also straightforward. The associated
Maxwell stress contributes to the Einstein equation without perturbing the
non-Riemannian fields. Indeed an Einstein-Weyl-Maxwell solution  exists
with Weyl form \gpole,\
Maxwell 1-form
$$A= \frac{q_0\,\dd t}{r}+\mu_0 \cos\theta \,\dd \phi\Eqno$$
with constants $q_0$ and $\mu_0$
and metric given by \metricsol\ where
$\kappa_2\,\mu^2\mapsto \kappa_2\,\mu^2+\mu_0^2$
and
$\kappa_2 \, q^2\mapsto \kappa_2 \,q^2+ q_0^2.$

The significance of non-Riemannian gravitational fields has long been
recognised by a number of coworkers. Their relevance in recent low energy
effective actions has stimulated a renewed interest in the physical
significance of these fields.
The solutions discussed above may offer an opportunity to confront theoretical
predictions with some of the classical tests of general relativity and place
bounds on the couplings that enter in the modifications to Einstein's
theory. They also raise intriguing questions about the nature of the
coupling of  test particles with Weyl charge
 to the non-Riemannian components of the
gravitational field.
The relevance of such non-Riemannian fields in the
astrophysical sector will be discussed elsewhere.

\Section{ Acknowledgments}

RWT is grateful to R Kerner for providing facilities at the
Laboratoire de Gravitation et Cosmologie Relativistes, Universite Pierre et
Marie Curie, CNRS,  Paris where this work was begun and
to the Human Capital and Mobility Programme of the European Union for
partial
support. CW is grateful
to the School of Physics and Chemistry,
University of Lancaster  for a School Studentship,
to the Committee of Vice-Chancellors and Principals,
UK for an Overseas Research
Studentship
and to the University of Lancaster for a Peel Studentship.
We are grateful to F Hehl for pointing out reference \added\ to us.

\vfill\eject

\References

\bye